\documentclass[12pt,a4paper]{article}

\usepackage{amsmath}
\usepackage{amssymb}
\usepackage{epsfig}
\usepackage{graphics}

\def\nostrocostrutto#1\over#2{\mathrel{\mathop{\kern 0pt \rlap 
  {\raise.2ex\hbox{$#1$}}}
  \lower.9ex\hbox{\kern-.190em $#2$}}}

\catcode`@=11
\newcount\@tempcntc
\def\@citex[#1]#2{\if@filesw\immediate\write\@auxout{\string\citation{#2}}\fi
  \@tempcnta\z@\@tempcntb\m@ne\def\@citea{}\@cite{\@for\@citeb:=#2\do
    {\@ifundefined
       {b@\@citeb}{\@citeo\@tempcntb\m@ne\@citea\def\@citea{,}{\bf ?}\@warning
       {Citation `\@citeb' on page \thepage \space undefined}}%
    {\setbox\z@\hbox{\global\@tempcntc0\csname b@\@citeb\endcsname\relax}%
     \ifnum\@tempcntc=\z@ \@citeo\@tempcntb\m@ne
       \@citea\def\@citea{,}\hbox{\csname b@\@citeb\endcsname}%
     \else
      \advance\@tempcntb\@ne
      \ifnum\@tempcntb=\@tempcntc
      \else\advance\@tempcntb\m@ne\@citeo
      \@tempcnta\@tempcntc\@tempcntb\@tempcntc\fi\fi}}\@citeo}{#1}}
\def\@citeo{\ifnum\@tempcnta>\@tempcntb\else\@citea\def\@citea{,}%
  \ifnum\@tempcnta=\@tempcntb\the\@tempcnta\else
   {\advance\@tempcnta\@ne\ifnum\@tempcnta=\@tempcntb \else \def\@citea{--}\fi
    \advance\@tempcnta\m@ne\the\@tempcnta\@citea\the\@tempcntb}\fi\fi}
\catcode`@=12

\begin{document}




\setcounter{page}{0}
\thispagestyle{empty}
\begin{titlepage}

\vspace*{-1cm}
\hfill \parbox{3.5cm}{ 
BUTP-2001/13
\vspace*{0.3cm} \\
17. May, 2001 
\vspace*{0.3cm}
 }   
\vfill

\begin{center}
  {\large {\bf
On simple and subtle properties of neutrinos
} 
      \footnote{We thank the Schweizerische Nationalfonds for his support.}  }
\vfill
\vspace*{0.3cm} 

{\bf
    Peter Minkowski } \\
    Institute for Theoretical Physics \\
    University of Bern \\
    CH - 3012 Bern, Switzerland
    \\
    E-mail: mink@itp.unibe.ch
   \vspace*{0.3cm} \\  

\end{center}

\vfill

\begin{abstract}
\noindent
Neutrino flavors, light and heavy, are discussed in their interdependence
within the minimal unifying gauge group of SO10. The general situation
which excludes the existence of an {\it exact} symmetry from
which {\it exactly} vanishing light neutrino masses would follow
is discussed. {\it Subtle} and {\it simple} consequences
for 'low-low' oscillation phenomena are presented in a general
framework.
\end{abstract}



\vfill
\end{titlepage}



\newpage


\section{SO10 extension of the standard model : \protect\\
\hspace*{3.5cm}$\left ( \ \nu \ , \ {\cal{N}} \ \right )^{ \dot{\gamma}}_{\ (J)}$}
\vspace*{0.1cm} 

\noindent
In the title above the label $ ^{ \dot{\gamma}}$ denotes, for $\dot{\gamma} \ = \ 1,2$,
the two left-chiral components of the spin 1/2 fields $\nu$ and ${\cal{N}}$ respectively.

\noindent
The label (J), for $J \ = \ 1,2,3$, denotes family number. So we define
the (minimal) fermionic extension of the three standard model
fermion families to three families of 16-representations of SO10 :

\begin{equation}
\label{eq:1}
\begin{array}{l}
\left \lbrace \ f \ \right \rbrace^{\ \dot{\gamma}}_{\ (J)}
\ = 
\left \lbrace 
\begin{array}{l}
\ u^{\ 1 \ , \ 2 \ , \ 3} \ , \ \nu_{\ e} \ \left | \hspace*{0.2cm} {\cal{N}}_{\ e} \ , 
\ \hat{u}^{\ 1 \ , \ 2 \ , \ 3} \right .
\vspace*{0.3cm} \\
\ d^{\ 1 \ , \ 2 \ , \ 3} \ , \ e^{\ -} \ \left | \hspace*{0.2cm} \hat{e}^{\ +} \ , 
\ \hat{d}^{\ 1 \ , \ 2 \ , \ 3} \right .
\end{array}
\ \right \rbrace^{\ \dot{\gamma}}_{\ (J)}
\end{array}
\end{equation}

\noindent
Under the standard model gauge group $SU3_{\ c} \ \times \ SU2_{\ L} \ \times \ U1_{\ {\cal{Y}}}$
the fermion flavors in eq. (\ref{eq:1}) transform as 

\begin{equation}
\label{eq:2}
\begin{array}{l}
\left \lbrace 
\begin{array}{l}
\ u^{\ 1 \ , \ 2 \ , \ 3} 
\vspace*{0.3cm} \\
\ d^{\ 1 \ , \ 2 \ , \ 3} 
\end{array}
\ \right \rbrace
\ = \ \left ( \ 3 \ , \ 2 \ , \ \frac{1}{6} \ \right )
\hspace*{0.1cm} ; \hspace*{0.1cm}
\left \lbrace 
\begin{array}{l}
\ \nu_{\ e}
\vspace*{0.3cm} \\
\ e^{\ -} 
\end{array}
\ \right \rbrace
\ = \ \left ( \ 1 \ , \ 2 \ , \ - \frac{1}{2} \ \right )
\vspace*{0.3cm} \\
{\cal{N}}_{\ e}  
\ = \ \left ( \ 1 \ , \ 1 \ , \ 0 \ \right )
\hspace*{0.1cm} ; \hspace*{0.1cm}
\hat{e}^{\ +} 
\ = \ \left ( \ 1 \ , \ 1 \ , \ 1 \ \right )
\vspace*{0.3cm} \\
\left \lbrace \ \hat{u}^{\ 1 \ , \ 2 \ , \ 3} \ \right \rbrace
\ = \ \left ( \ \overline{3} \ , \ 1 \ , \ - \frac{2}{3} \ \right )
\hspace*{0.1cm} ; \hspace*{0.1cm}
\left \lbrace \ \hat{d}^{\ 1 \ , \ 2 \ , \ 3} \ \right \rbrace
\ = \ \left ( \ \overline{3} \ , \ 1 \ , \ \frac{1}{3} \ \right )
\end{array}
\end{equation}

\noindent
We ask as a guiding question whether the associated B - L (baryon number - lepton number)
charge with quantum numbers

\begin{equation}
\label{eq:3}
\begin{array}{l}
\left \lbrace 
\begin{array}{l}
\ \frac{1}{3} \ , \ \frac{1}{3} \ , \ \frac{1}{3} \ , \ - 1
\left | \hspace*{0.2cm} + 1 \ , 
\ - \frac{1}{3} \ , \ - \frac{1}{3} \ , \ - \frac{1}{3}
\ \right .
\vspace*{0.3cm} \\
\ \frac{1}{3} \ , \ \frac{1}{3} \ , \ \frac{1}{3} \ , \ - 1
\left | \hspace*{0.2cm} + 1 \ , 
\ - \frac{1}{3} \ , \ - \frac{1}{3} \ , \ - \frac{1}{3} 
\ \right .
\end{array}
\ \right \rbrace
\end{array}
\end{equation}

\noindent
can be conserved - if not gauged - in the limit 
$m_{\ {\cal{N}} \ (J)} \ \rightarrow \ \infty$.

\noindent
The limit of infinite mass for the (standard model gauge group) singlet neutrino flavors 
${\cal{N}}_{\ e \ (j)}$ implies a Majorana mass matrix of the form  

\begin{equation}
\label{eq:4}
\begin{array}{l}
{\cal{H}}_{\ m} \ = \ \frac{1}{2} 
\ m_{\ {\cal{N}} \ J \ J'} 
\ {\cal{N}}_{\ \dot{\gamma} \ J}  
\ {\cal{N}}^{\ \dot{\gamma}}_{\ J'}  
\ + \ h.c.
\vspace*{0.3cm} \\
m_{\ {\cal{N}} \ J \ J'} \ = \ m_{\ {\cal{N}} \ J' \ J} \ \rightarrow \ M_{\ J \ J'}
\hspace*{0.2cm} ; \hspace*{0.2cm}
M \ \rightarrow \ \infty
\end{array}
\end{equation}

\noindent
Apart from beeing symmetric the mass matrix $m_{\ {\cal{N}}}$
in eq. (\ref{eq:4}) is complex general and the physical masses in the
infinite mass limit involve the eigenvalues of the associated hermitian matrices
$( \ m \ m^{\ \dagger} \ )^{\ 1/2}$ and (equivalently)
$( \ m^{\ \dagger} \ m \ )^{\ 1/2}$.

\noindent
Including the ${\cal{N}}_{\ (J)}$ flavors (before taking the infinite mass limit)
the B - L current is of vectorial form. In a chiral basis this amounts to a
matching number of positive and negative eigenvalues of the B - L charge as exhibited
in eq. (\ref{eq:3}) :

\begin{equation}
\label{eq:5}
\begin{array}{l}
j_{\ \mu}^{\ B - L \ (16)} \ = \ \sum_{\ J}
\ \sum_{\ r}^{\ 16} \ ( \ B \ - \ L \ )_{\ r}
\ f^{\ * \ \beta}_{\ r \ (J)} \ \sigma_{\ \mu \ \beta \ \dot{\gamma}} 
\ f^{\ \dot{\gamma}}_{\ r \ (J)}
\end{array}
\end{equation}

\noindent
For finite masses $m_{\ {\cal{N}} \ J}$ the B - L current is broken by these masses,
but in the infinite mass limit the remaining (45) flavors are constrained 
to form a chiral current :

\begin{equation}
\label{eq:6}
\begin{array}{l} 
j_{\ \mu}^{\ B - L \ (15)} \ = \ \sum_{\ J}
\ \sum_{\ r}^{\ 15} \ ( \ B \ - \ L \ )_{\ r}
\ f^{\ * \ \beta}_{\ r \ (J)} \ \sigma_{\ \mu \ \beta \ \dot{\gamma}} 
\ f^{\ \dot{\gamma}}_{\ r \ (J)}
\end{array}
\end{equation}

\noindent
This reduced chiral current ($j_{\ \mu}^{\ B - L \ (15)}$ in eq. (\ref{eq:6}))
develops a gravitational anomaly and thus fails to be conserved in a gravitational environment :

\begin{equation}
\label{eq:7}
\begin{array}{l} 
D^{\ \mu} \ j_{\ \mu}^{\ B - L \ (15)} \ = 
\ 3 \ c_{\ 1} \ ( \ spin \ ) \ p_{\ 1} \ ( \ R \ )
\ = \ 3 \ {\cal{A}}_{\ 1} \ ( \ R \ )
\vspace*{0.3cm} \\
c_{\ 1} \ ( \ spin \ ) \ = \ - \ \frac{1}{24}
\end{array}
\end{equation}

\noindent
In eq. (\ref{eq:7}) $D_{\ \mu}$ denotes the covariant derivative with respect to the vierbein.
Since the current is equivalent to an antisymmetric 3-form (in four dimensions) 
the divergence does not involve the vierbein or the metric and thus the right hand side
of eq. (\ref{eq:7}) necessarily defines a topological 4-form.

\noindent
This 4-form defines, apart from the overall factor 3, the Hirzebruch-Atiah index form. 
The integral of this form yields the chiral topological invariant of the
(full) Dirac operator pertaining to the curvature 2-form of the metric (or vierbein).

\noindent
It is also referred to as ${\cal{A}}$ genus in the (mathematical) literature
and extends to any (Euclidean (compact)) space with dimension divisable by 4 :
$d_{\ n} \ = \ 4 \ n$.

\noindent
The coefficient $c_{\ n} \ ( \ spin  \ )$ gives the relation between
the Hirzebruch-Atiah- and Pontrjagin classes in 
$d_{\ n} \ = \ 4 \ n$ dimensions. The latter are generically (for all n simultaneously)
defined through the relations

\begin{equation}
\label{eq:8}
\begin{array}{l} 
\overline{R}^{\ a}_{\ b} \ = 
\ \frac{1}{2 \ \pi} \ \frac{1}{2}
\ d \ x^{\ \mu} \ \wedge \ d \ x^{\ \nu} 
\ \left ( \ R^{\ a}_{\ b} \ \right )_{\ \mu \nu}
\vspace*{0.3cm} \\
Det \ ( \ 1 \ - \ \lambda \ \overline{R} \ )
\ = \ \sum_{\ n} \ \lambda^{\ 2n} \ p_{\ n} \ ( \ R \ )
\vspace*{0.3cm} \\
p_{\ 1} \ = \ \frac{1}{16 \ \pi^{\ 2}} \ R^{\ a}_{\ b \ \mu \ \nu}
\ \widetilde{R}^{\ b \ \mu \ \nu}_{\ a}
\hspace*{0.2cm} ; \hspace*{0.2cm}
\widetilde{R}^{\ b \ \mu \ \nu}_{\ a} \ =
\ \frac{1}{2} \ \varepsilon^{\ \mu \nu \sigma \tau}
\ \left ( \ R^{\ a}_{\ b} \ \right )_{\ \sigma \tau}
\vspace*{0.3cm} \\
p_{\ 2} \ = \ - \ \frac{1}{4} \ tr \ \overline{R}^{\ 4}
\ + \ \frac{1}{8} \ \left ( \ tr \ \overline{R}^{\ 2} \ \right )^{\ 2}
\hspace*{0.2cm} ; \hspace*{0.2cm}
\cdots
\end{array}
\end{equation}

\noindent
with

\begin{equation}
\label{eq:9}
\begin{array}{l} 
{\cal{A}} \ ( \ R \ ; \ \lambda \ )
\ = \ Det 
\ \left (
\ \begin{array}{c}
\lambda \ \frac{1}{2} \ \overline{R}
 \vspace*{0.3cm} \\
\hline	\vspace*{-0.3cm} \\
\sin 
\ \left ( \ \lambda \ \frac{1}{2} \ \overline{R} \ \right )
\end{array}
\ \right )
\ = \ \sum_{\ n} \ \lambda^{\ 2 n} \ c_{\ n} \ ( \ R \ )
\vspace*{0.3cm} \\
c_{\ 1} \ = \ - \ \frac{1}{24} \ p_{\ 1}
\hspace*{0.2cm} ; \hspace*{0.2cm}
c_{\ 2} \ = \ 
\ \begin{array}{c}
1
 \vspace*{0.3cm} \\
\hline	\vspace*{-0.3cm} \\
2^{\ 7} \ 3^{\ 2} \ 5
\end{array}
\ \left (
\ - \ 4 \ p_{\ 2} \ + \ 7 \ p_{\ 1}^{\ 2}
\ \right )
\hspace*{0.2cm} ; \hspace*{0.2cm}
\cdots
\end{array}
\end{equation}

\noindent
- the classes (Hirzebruch-Atiyah and Pontrjagin classes) are {\it simple}.

\noindent
- the relations are {\it obvious} but not clearly simple.

\noindent
- the consequence is {\it subtle} : There does not exist any exact symmetry, which 
\hspace*{0.3cm} is {\it necessary} to maintain exactly massless neutrino flavors. 

\noindent
\hspace*{0.3cm} This is so even if the overall (6 chiral flavor) neutrino mass matrix

\noindent
\hspace*{0.3cm} would exactly conserve B - L.

\noindent
In order to have exact B - L symmetry the six chiral neutrino flavors have to combine
into triply doubled (Dirac) {\it pairs} with equal overall B - L quantum numbers :

\begin{equation}
\label{eq:10}
\begin{array}{l} 
\left \lbrace \ \nu_{\ A} \ \right \rbrace_{\ (J)} \ =
\ \left ( 
\ \begin{array}{l} 
 \varepsilon_{\ \alpha \beta} \ ( \ {\cal{N}}^{\ \dot{\beta}} \ )^{\ * \ (J)}
\vspace*{0.3cm} \\ 
\nu^{\ \dot{\gamma}}_{\ (J)}
\end{array}
\ \right )
\hspace*{0.2cm} ; \hspace*{0.2cm}
A \ = \ 1,\cdots,4
\vspace*{0.3cm} \\ 
A \ = \ 1,2 \ \leftrightarrow \ \alpha \ = \ 1,2
\hspace*{0.2cm} ; \hspace*{0.2cm}
A \ = \ 3,4 \ \leftrightarrow \ \dot{\gamma} \ = \ 1,2
\end{array}
\end{equation}

\noindent
with a B - L conserving mass matrix of the form equivalent to the case of 
(electrically) charged flavors :

\begin{equation}
\label{eq:11}
\begin{array}{l} 
{\cal{H}}_{\ m}^{\ \nu , {\cal{N}}} \ =
m^{\ \nu , \ {\cal{N}}}_{\ J \ J'} \ \overline{\nu}_{\ J} 
\ \frac{1}{2} \ ( \ 1 \ + \ \gamma_{\ 5 \ L} \ ) 
\ \nu_{\ J'} \ + \ h.c.
\vspace*{0.3cm} \\ 
\gamma_{\ 5 \ L} \ = 
\ - \ \gamma_{\ 5 \ R} \ = 
i \ \gamma_{\ 0} \ \gamma_{\ 1} \ \gamma_{\ 2} \ \gamma_{\ 3}
\end{array}
\end{equation}

\noindent
The reduced mass matrix $m^{\ \nu , {\cal{N}}}_{\ J \ J'}$ is an unrestricted complex
3 by 3 matrix. 

\noindent
The physical masses are the eigenvalues of the hermitian (nonnegative) combination
$ \left ( \ m^{\ \nu , \ {\cal{N}}} m^{\ \dagger \ \nu , \ {\cal{N}}} \ \right )^{\ 1/2}$.

\noindent
While in the above situation B - L is exactly conserved and therefore not gauged,
three arbitrary neutrino masses are allowed.

\noindent
Using the (Dirac doubled) field variables $\left \lbrace \ \nu_{\ A} \ \right \rbrace_{\ (J)}$
as defined in eqs. (\ref{eq:10}) and (\ref{eq:11}) the conserved B - L
current is given by

\begin{equation}
\label{eq:12}
\begin{array}{l} 
j_{\ \mu}^{\ B - L \ (16)} \ = \ \sum_{\ J}
\ \left (
\ \begin{array}{l} 
\ \frac{1}{3} \ \sum_{\ q = u , d \ , \ c} \ \overline{q}^{ c}_{\ (J)} 
\ \gamma_{\ \mu} \ q^{\ c}_{\ (J)}
\vspace*{0.3cm} \\ 
 - \ \overline{\nu}_{\ (J)}  
\ \gamma_{\ \mu} \ \nu_{\ (J)}
\vspace*{0.3cm} \\ 
 - \ \overline{e}_{\ (J)} \ \gamma_{\ \mu} \ e_{\ (J)} 
\end{array}
\ \right )
\end{array}
\end{equation}

\noindent
A {\it simpler} form of the {\it subtle} consequence just outlined is :

\noindent
there is no neutrino flavor with exactly vanishing mass \cite{Brems}. 
\footnote{The diploma thesis cited contains common work with P.M.}
\vspace*{0.2cm}  

\noindent
It is mandatory to mention here, which type of sometimes tacitly assumed consequence(s)
do {\it not} follow from this consequence but rather constitute just a logical
{\it possibility} :

\noindent
- neutrino flavors come in (Dirac) doubled ( B - L ) conserving and mass 

\noindent
\hspace*{0.3cm} degenerate pairs. 

\noindent
- the mixing matrix of the observed three light neutrino flavors is 

\noindent
\hspace*{0.3cm} exactly unitary.

\section{The actual SO10 scenario}
\vspace*{0.1cm}  

\noindent
Any extension of the standard model, which on the level of the gauge group
contains SO10 (irrespective of whether this extension contains supersymmetry or not)
leads to several {\it simple} consequences :

\noindent
- B - L as well as all other leptonic numbers are broken (spontaneously).

\noindent
- the minimum of neutrino flavors is six, and any mass degeneracy of these {\it six}
  flavors is {\it accidental}.

\noindent
- the mixing matrix of any three out of the six neutrino flavors is {\it not} unitary.

\noindent
- the number of real CP violating parameters, associated {\it just} with the mass

\noindent
\hspace*{0.3cm} matrix of the minimally six neutrino flavors is 15. These are all observable,

\noindent
\hspace*{0.3cm} albeit with variable sensitivity (at relatively) low energies.
\vspace*{0.1cm}  

\noindent
{\bf A simple phenomenological excursion}
\vspace*{0.1cm}  

\noindent
Let us 'fix ideas' with respect to the three light neutrino flavors, and
restricting the {\it assumed} heavy ones to three, {\it assuming}
hierarchical masses \cite{Petcov}
\footnote{The cited paper gives an extensive phenomenological review.} :

\begin{equation}
\label{eq:14}
\begin{array}{l} 
\left \lbrace \ m^{\ \nu,{\cal{N}}} \ \right \rbrace
\ = \ \left \lbrace \ m_{\ 1,2,3} \ , \ M_{\ 1,2,3} \ \right \rbrace
\vspace*{0.3cm} \\ 
M_{\ 1,2,3} \ \gg \ 1 \ \mbox{GeV}
\hspace*{0.2cm} ; \hspace*{0.2cm} 
m_{\ 1,2,3} \ \ll \ 1 \ \mbox{GeV}
\vspace*{0.3cm} \\ 
m_{\ 3} \ \gg \ m_{\ 2} \ > \ \mbox{or} \ \gg \ m_{\ 1}
\hspace*{0.2cm} ; \hspace*{0.2cm} 
\Delta_{\ i j}^{\ 2} \ = \ m_{\ i}^{\ 2} \ - \ m_{\ j}^{\ 2}
\vspace*{0.3cm} \\ 
\Delta_{\ 3 2}^{\ 2} \ = \ 3. \ 10^{\ -3} \ eV^{\ 2}
\hspace*{0.2cm} ; \hspace*{0.2cm} 
\mbox{to be modified eventually}
\vspace*{0.3cm} \\ 
\Delta_{\ 3 2} \ = \ 0.055 \ eV \ = \ 632 \ ^{o} \ K
\vspace*{0.3cm} \\ 
\Delta_{\ 2 1}^{\ 2} \ = \ 3.5 \ 10^{\ -5} \ eV^{\ 2} 
\hspace*{0.2cm} ; \hspace*{0.2cm} 
\mbox{to be modified eventually}
\vspace*{0.3cm} \\ 
\Delta_{\ 2 1} \ = \ 0.0059 \ eV \ = \ 68.5 \ ^{o} \ K
\end{array}
\end{equation}

\noindent
This yields, reemphasizing the elimination {\it by assumption} 
of a common mass for the three light neutrino flavors much larger than the mass differences

\begin{equation}
\label{eq:15}
\begin{array}{l} 
m_{\ 3} \ \simeq \ 0.055 \ eV \ = \ 632 \ ^o \ K
\vspace*{0.3cm} \\ 
m_{\ 2} \ \simeq \ 0.0059 \ eV \ = \ 68.5 \ ^o \ K
\vspace*{0.3cm} \\ 
m_{\ 1} \hspace*{3.0cm} \stackrel{>}{<} \ 2 \ ^o \ K
\end{array}
\end{equation}

\noindent
Then the mass eigenstates 3 and 2 constitute hot dark matter at the time of nucleosynthesis
but cold today, whereas the lightest neutrino flavor 1 may be still in relativistic (mean)
motion today.

\noindent
The general $( \ \nu \ , \ {\cal{N}} \ )$ mass term is of the form

\begin{equation}
\label{eq:16}
\begin{array}{c} 
{\cal{H}}_{\ m}^{\ \nu , {\cal{N}}} \ =
\ \frac{1}{2} \ \nu^{\ \dot{\delta}}_{\ i} \ \varepsilon_{\ \dot{\gamma} \dot{\delta}}
\ {\cal{M}}_{\ i k} \ \nu^{\ \dot{\gamma}}_{\ k} \ + \ h.c.
\hspace*{0.2cm} ; \hspace*{0.2cm} 
{\cal{M}}_{\ i k} \ = \ {\cal{M}}_{\ k i}
\vspace*{0.3cm} \\ 
i,k \ = \ 1, \cdots, \ 3 \ + \ N_{\ h}
\hspace*{0.2cm} ; \hspace*{0.1cm} 
r \ = \ k \ - \ 3 \ = \ 1, \cdots, N_{\ h}
\hspace*{0.1cm} ; \hspace*{0.1cm} 
\varepsilon_{\ \dot{\gamma} \dot{\delta}} \ {\cal{N}}^{\ \dot{\delta}}_{\ r}
\ = \ {\cal{N}}_{\ \dot{\gamma} \ r}
\vspace*{0.3cm} \\ 
{\cal{M}} \ =
\ \left (
\ \begin{array}{ll} 
0 & \ \mu^{\ T}
\vspace*{0.3cm} \\ 
\mu & M
\end{array}
\ \right )
\end{array}
\end{equation}

\noindent
In eq. (\ref{eq:16}) we generalize to $N_{\ h} \ \geq \ 3$ heavy neutrino flavors,
while the minimal SO10 scenario has $N_{\ h} \ = \ 3$.

\noindent
The mass matrix ${\cal{M}}$ is compatible with the so extended standard model
renormalizability conditions, provided the $N_{\ h} \ \times \ 3$ matrix $\mu$
and its transpose are generated through Yukawa couplings of one or several 
electroweak scalar doublets.

\begin{equation}
\label{eq:17}
\begin{array}{l} 
- \ {\cal{L}}_{\ Y} \ =
\ g_{\ k i}^{\ (\alpha)} 
\ {\cal{N}}_{\ \dot{\gamma} \ k - 3}
\ \left ( \ \phi^{\ 0}_{\ (\alpha)} \ , \ \phi^{\ -}_{\ (\alpha)} 
\ \right )^{\ *} 
\ \left (
\ \begin{array}{l}
\nu^{\ i \ \dot{\gamma}}
\vspace*{0.3cm} \\ 
e^{ - \ i \ \dot{\gamma}}
\end{array}
\ \right )
\hspace*{0.2cm} + \ h.c. 
\vspace*{0.3cm} \\ 
i \ = \ 1,2,3 \hspace*{0.2cm} ; \hspace*{0.2cm} k \ = \ 4, \cdots, \ 3 \ + \ N_{\ h}
\hspace*{0.2cm} ; \hspace*{0.2cm} 
\alpha \ = \ 1, \cdots, \ n_{\ sc}
\vspace*{0.3cm} \\ 
\mu_{\ k i} \ = \ g_{\ k i}^{\ (\alpha)} \ v^{\ ch}_{\ (\alpha)}
\hspace*{0.2cm} ; \hspace*{0.2cm} 
v^{\ ch}_{\ (\alpha)} \ = 
\ \left \langle \ 
\ \Omega \ \right |
\ \phi^{\ 0}_{\ (\alpha)}  
\ \left | \ \Omega \ \right \rangle^{\ *}
\end{array}
\end{equation}

\noindent
In eq. (\ref{eq:17}) $\alpha$ numbers scalar doublets. For one doublet in the SM
or two doublets in the MSSM the vacuum expected values are 

\begin{equation}
\label{eq:18}
\begin{array}{l} 
SM \hspace*{1.0cm} : \ n_{\ sc} \ = \ 1
\hspace*{0.2cm} ; \hspace*{0.2cm} 
v^{\ ch}_{\ (1)} \ \rightarrow \ \frac{1}{\sqrt{2}} \ v 
\hspace*{0.2cm} ; \hspace*{0.2cm} 
v \ = \ \left ( \ G_{\ F} \ \sqrt{2} \ \right )^{\ -1/2}
\vspace*{0.4cm} \\ 
MSSM \ : \ n_{\ sc} \ = \ 2
\hspace*{0.2cm} ; \hspace*{0.2cm} 
\left (
\begin{array}{l} 
v^{\ ch}_{\ (1)} \ \rightarrow \ \frac{1}{\sqrt{2}} \ v^{\ u} 
\vspace*{0.3cm} \\ 
v^{\ ch}_{\ (2)} \ \rightarrow \ 0 
\end{array}
\right )
\hspace*{0.2cm} ; \hspace*{0.2cm} 
v^{\ \ u} \ = \ \sin \ \beta \ v
\end{array}
\end{equation}

\noindent
{\bf Generic estimate of light versus heavy flavors}
\vspace*{0.1cm}  

\noindent
A {\it simple} consequence of the structure of ${\cal{M}}$ as defined in eq. (\ref{eq:16})
concerns the absolute value of the determinant, given the {\it assumption}
that each (complex) eigenvalue of the $N_{\ h} \ \times \ N_{\ h}$ matrix M
is much larger in absolute value than (the absolute value of) any given element 
of the $N_{\ h} \ \times \ 3$ matrix $\mu$.

\noindent
It then follows for the (complex) determinant of ${\cal{M}}$ :

\begin{equation}
\label{eq:19}
\begin{array}{l} 
Det \ {\cal{M}} \ = \ Det \ {\cal{M}}^{\ '}
\hspace*{0.2cm} ; \hspace*{0.2cm} 
{\cal{M}}^{\ '} \ =
\ \left (
\ \begin{array}{ll} 
0 & \ \mu^{\ ' \ T}
\vspace*{0.3cm} \\ 
\mu^{\ '} & M^{\ '}
\end{array}
\ \right )
\vspace*{0.4cm} \\ 
\mu^{\ '} \ =
\ \left (
\ \begin{array}{l|@{\hspace*{0.2cm}}lll} 
k \ = \ 4 & \hat{\mu}_{\ 11} & \hat{\mu}_{\ 12} &  \hat{\mu}_{\ 13} \\ 
k \ = \ 5 & \hat{\mu}_{\ 21} & \hat{\mu}_{\ 22} &  \hat{\mu}_{\ 23} \\ 
k \ = \ 6 & \hat{\mu}_{\ 31} & \hat{\mu}_{\ 32} &  \hat{\mu}_{\ 33} \\ 
k \ = \ 7 & \multicolumn{3}{c} {0} \\ 
\cdots    & \multicolumn{3}{c} {0} \\ 
k \ = \ N_{\ h} & \multicolumn{3}{c} {0} 
\end{array}
\ \right )
\vspace*{0.4cm} \\ 
M^{\ '} \ =
\ \left (
\ \begin{array}{cc} 
0_{\ 3 \times 3} & 0_{\ 3 \times 3} 
\vspace*{0.3cm} \\ 
0_{\ 3 \times 3} & \hat{M} 
\end{array}
\ \right )
\hspace*{0.2cm} ; \hspace*{0.2cm} 
\hat{M} \ = \hat{M}_{\ \kappa \lambda}
\vspace*{0.4cm} \\ 
k,l \ = \ \kappa \ + \ 3,\lambda \ + \ 3 \ = \ 7, \ \cdots, \ N_{\ h}
\end{array}
\end{equation}

\noindent
In eq. (\ref{eq:19}) $\hat{\mu}$ is formed from any subset of three out of 
the $N_{\ h}$ (three component -) row vectors of $\mu$ with maximal rank, which we
generically assume to be maximal, i.e. three.

\noindent
For the case of minimal heavy neutrino flavors $N_{\ h} \ = \ 3$,
we have {\it simply} $\hat{\mu} \ = \ \mu$ and $\hat{M} \ \rightarrow \ 1$.
\vspace*{0.2cm} 

\noindent
Modulo {\it simple} permutations according to the subset of three row vectors in $\mu$
chosen to form $\hat{\mu}$, the $N_{\ h} \ - \ 3 \ \times \ N_{\ h} \ - \ 3$
matrix $\hat{M}$ is of the form

\begin{equation}
\label{eq:20}
\begin{array}{l} 
\hat{M}_{\ \kappa \lambda} \ = 
\ \left [
\begin{array}{l} 
 M_{\ \kappa \lambda} \ + 
\ \left ( \ L_{\ \kappa}^{\ (i)} \ M_{\ (i) \lambda} \ + 
\ \kappa \ \leftrightarrow \ \lambda
\ \right )
\vspace*{0.3cm} \\ 
 + \ L_{\ \kappa}^{\ (i)} \ M_{\ (i) (j)} \ L_{\ \lambda}^{\ (j)}
\end{array}
\ \right ]
\vspace*{0.3cm} \\ 
| \ L_{\ \kappa}^{\ (i)} \ | \ = \ O \ (1)
\hspace*{0.2cm} ; \hspace*{0.2cm} 
(i),(j) \ = \ 1,2,3
\end{array}
\end{equation}

\noindent
Except for {\it non-generic} matrices $\mu$ and $M$ the eigenvalues of $\hat{M}$
($N_{\ h} \ - 3$ in number) are of the same order of magnitude than those
of ${\cal{M}}$ ($N_{\ h}$ in number).

\newpage

\noindent
From the structure of ${\cal{M}}^{\ '}$ in eq. (\ref{eq:19}) it follows

\begin{equation}
\label{eq:21}
\begin{array}{l} 
| \ Det \ {\cal{M}} \ | \ = 
\ ( \ \prod_{\ i}^{\ 3} \ m_{\ i} \ )  
\ ( \ \prod_{\ J}^{\ N_{\ h}} \ M_{\ J} \ )  
\ = 
\vspace*{0.3cm} \\ 
\ | \ Det_{\ 3 \times 3} \ \hat{\mu} \ |^{\ 2}  
\ | \ Det_{\ N_{\ h} - 3 \times N_{\ h} - 3} \ \hat{M} \ |
\end{array}
\end{equation}

\noindent
In eq. (\ref{eq:21}) (extending eq. (\ref{eq:14})) $M_{\ 1} \ , \cdots, \ M_{\ N_{\ h}}$
denote the masses of the heavy flavors.

\noindent
The above {\it simple} relations are known as 'sea-saw' mechanism \cite{MGMYan}.
\vspace*{0.1cm} 

\noindent
Introducing the geometric mean (nonnegtive) masses

\begin{equation}
\label{eq:22}
\begin{array}{l} 
\prod_{\ i}^{\ 3} \ m_{\ i} \ = \ m_{\ light}^{\ 3}  
\hspace*{0.2cm} ; \hspace*{0.2cm} 
\prod_{\ J}^{\ N_{\ h}} \ M_{\ J} \ = \ M_{\ heavy}^{\ N_{\ h}}  
\vspace*{0.3cm} \\ 
| \ Det_{\ 3 \times 3} \ \hat{\mu} \ | \ = \ \overline{\mu}^{\ 3} 
\hspace*{0.2cm} ; \hspace*{0.2cm} 
| \ Det_{\ N_{\ h} - 3 \times N_{\ h} - 3} \ \hat{M} \ |
\ = \ \overline{M}^{\ N_{\ h} \ - \ 3} 
\end{array}
\end{equation}

\noindent
the relation in eq.(\ref{eq:21}) takes the form

\begin{equation}
\label{eq:23}
\begin{array}{l} 
m_{\ light} \  M_{\ heavy}  
\ =
\ \overline{\mu}^{\ 2} 
\ \left (
\ \begin{array}{c}
\overline{M}
 \vspace*{0.3cm} \\
\hline	\vspace*{-0.3cm} \\
M_{\ heavy}
\end{array}
\ \right )^{\ N_{\ h}/3 \ - \ 1} 
\end{array}
\end{equation}

\noindent
The estimate of the Yukawa induced (doublet-singlet) mass $\overline{\mu}$
is derived from the {\it simple} 16 $\times$ 16 $\times$ 10 \hspace*{0.2cm} SO10 mass relation 
evoluted for quark flavors to a unification mass $M_{\ unif} \ \sim \ 10^{\ 16} \ GeV$ :

\begin{equation}
\label{eq:24}
\begin{array}{l} 
\overline{\mu} \ = \ C_{\ \nu} \ \frac{1}{3} 
\ ( \ m_{\ u} \ m_{\ c} \ m_{\ t} \ )^{\ 1/3} \ \sim \ C_{\ \nu} \ 0.35 \ GeV
\hspace*{0.2cm} ; \hspace*{0.2cm} 
C_{\ \nu} \ = \ O \ (1)
\end{array}
\end{equation}

\noindent
Let us measure the geometric mean light neutrino mass $m_{\ light}$
in units of $10^{\ -2} \ eV$. Then the estimate in eq. (\ref{eq:23}) takes the form 

\begin{equation}
\label{eq:25}
\begin{array}{l} 
m_{\ light} \ = \ K_{\ light} \ 10^{\ -11} \ GeV
\hspace*{0.2cm} \rightarrow \hspace*{0.2cm} 
 \vspace*{0.3cm} \\
M_{\ heavy}  
\ \sim \ 1.2 \ 10^{\ 10} \ GeV
\ ( \ C_{\ \nu} / K_{\ light} \ )
\ \left (
\ \begin{array}{c}
\overline{M}
 \vspace*{0.3cm} \\
\hline	\vspace*{-0.3cm} \\
M_{\ heavy}
\end{array}
\ \right )^{\ N_{\ h}/3 \ - \ 1} 
\end{array}
\end{equation}

\noindent
For given light neutrino mass (in geometric mean) and given $\overline{\mu}$,
$M_{\ heavy}$ can be reduced through the ratio $\overline{M} / M_{\ heavy}$
even considerably if there are many heavy neutrino flavors beyond the minimal
three. However this emerges as the {\it simple} consequence of the light
neutrino flavors mixing predominantly to the heaviest three, while only
little to the {\it lighter} $N_{\ h} - 3$ ones.

\noindent
I wish to emphasize here, that any mass relations inside and outside of 
a unifying gauge group out of the increasing sequence 
$SO10 \ < \ E6 \ < ... < \ E8^{\ p}$ cannot
explain the pattern of light {\it and} heavy neutrino flavors. E.g. the mass matrix
M, remains totally unconstrained within the SM, yet within SO10 it has the
quantum numbers of the 126 irreducible representation. If there exist elementary scalars
transforming as this 126 (complex) representation the induced structure of M is intrinsically
tied to rest symmeties remaining after the breakdown of SO10 gauge invariance, 
beyond the symmetries of the SM. If these symmetries are associated with an N=1 or 2 susy 
structure again no clear mass relations among the known fermion families arise naturally.

\noindent
On the other hand the mass matrix M can be induced through finite loop effects involving
e.g. the square of scalar 16 representations, but again no direct structure reflecting
this situation on the known fermion families can be derived.
\vspace*{0.1cm}

\noindent
On the other hand the generic ratio

\begin{equation}
\label{eq:26}
\begin{array}{l} 
m_{\ light} \ / \ M_{\ heavy} \ \sim \ 10^{\ -21}
\hspace*{0.2cm} \leftrightarrow \hspace*{0.2cm} 
M_{\ heavy} \ \sim \ 10^{\ 7} \ TeV
\end{array}
\end{equation}

\noindent
is a {\it subtle} measure of lepton number violation at electroweak and lower energies,
as well as of associated CP violation. The generic mass scale $M_{\ heavy}$ is well
above the asssumed susy scale - by seven orders of magnitude - if the latter is assumed to be 1 TeV.
From this a dangereous enhancement of both CP-violation beyond the CKM matrix and
lepton number violation by these seven orders of magnitude indirectly
affect susy induced contributions to all electric dipole moments ( transition dipole
moments for neutrinos ) including the neutron, and directly the charged leptons,
as well as to lepton number violating processe
like $\mu \ \rightarrow \ e \ \gamma$ and $\mu \ \rightarrow 3 \ e$.

\noindent
This in my opinion augments the necessity to look for a theoretically viable explanation
of mass and mixing structures but it must be sought along hitherto unexplored paths.
Further new phenomena beyond the SM hopefully will give some clues.

\section{The Majorana equations for $\nu \ , \ {\cal{N}}$ flavors }
\vspace*{0.1cm} 

\noindent
In the following we restrict the discussion to three heavy neutrino flavors for simplicity.
Then the field equations for the associated chiral fields \\
$\nu^{\ \dot{\gamma}}_{\ j} \ , \ \nu^{\ *}_{\ \alpha \ j}$ j = 1,...,6 take
the form

\begin{equation}
\label{eq:27}
\begin{array}{l} 
\nu^{\ \dot{\gamma}}_{\ j+3} \ = {\cal{N}}^{\ \dot{\gamma}}_{\ j}
\hspace*{0.2cm} ; \hspace*{0.2cm} 
j \ = \ 1,2,3
 \vspace*{0.3cm} \\
\nu^{\ *}_{\ \alpha \ j} \ = \ \varepsilon_{\ \alpha \beta} 
\ \left (
\ \nu^{\ \dot{\beta}}_{\ j}
\ \right )^{\ *}
 \vspace*{0.3cm} \\
i \ D^{\ \dot{\gamma} \alpha} \ \nu^{\ *}_{\ \alpha} \ = \ {\cal{M}} \ \nu^{\ \dot{\gamma}}
 \vspace*{0.3cm} \\
i \ D_{\ \alpha \dot{\gamma}} \ \nu^{\ \dot{\gamma}} \ = \ {\cal{M}}^{\ \dagger} 
\ \nu^{\ *}_{\ \alpha}
 \vspace*{0.3cm} \\
 {\cal{M}}^{\ \dagger} \ = \ \overline{{\cal{M}}}
\hspace*{0.2cm} ; \hspace*{0.2cm} 
{\cal{M}} \ = \ {\cal{M}}^{\ T} 
\end{array}
\end{equation}

\noindent
In eq. (\ref{eq:27}) the covariant derivatives are restricted to gravity.

\noindent
{\bf Diagonalization to mass eigenstates $\nu \ \rightarrow \ \hat{\nu}$} 

\noindent
It is {\it simple} but somewhat involved \cite{PM} to unitarily diagonalize
the 6 by 6 matrix ${\cal{M}}$

\begin{equation}
\label{eq:28}
\begin{array}{l} 
{\cal{M}} \ = \ U \ {\cal{M}}_{\ D} \ U^{\ T}
\hspace*{0.2cm} ; \hspace*{0.2cm} 
U \ =
\ \left (
\begin{array}{ll} 
U_{\ (11)} & U_{\ (12)} 
 \vspace*{0.3cm} \\
U_{\ (21)} & U_{\ (22)} 
\end{array}
\ \right )
 \vspace*{0.3cm} \\
 {\cal{M}}_{\ D} \ = \ diag \ ( \ m_{\ 1,2,3} \ ; M_{\ 1,2,3} \ )
\end{array}
\end{equation}

\noindent
In eq. (\ref{eq:28}) we break up the 6 by 6 matrix U into 3 by 3 blocks,
since neutrino's prepared in a system which does not allow for the production of
heavy neutrino flavors involves only the {\it mass eigenfields} 
$\hat{\nu}_{\ k} \ ; \ k=1,2,3$, whereas again at electroweak energy scales
only (predominantly) the modes of $\nu_{\ j} \ ; \ j=1,2,3$ are produced 
through the electroweak interactions and are inducing a reaction
downstream the oscillation path.  We refer to this as  'low-low' oscillation physics.

\noindent
Eq. (\ref{eq:27}) becomes 

\begin{equation}
\label{eq:29}
\begin{array}{l} 
i \ D^{\ \dot{\gamma} \alpha} \ \left ( \ U^{\ T} \ \nu \ \right )^{\ *}_{\ \alpha} 
\ = \ {\cal{M}}_{\ D} \ \left ( \ U^{\ T} \ \nu \ \right )^{\ \dot{\gamma}}
 \vspace*{0.3cm} \\
i \ D_{\ \alpha \dot{\gamma}} \ \left ( \ U^{\ T} \ \nu \ \right )^{\ \dot{\gamma}} 
\ = \ {\cal{M}}_{\ D} \ \left ( \ U^{\ T} \ \nu \ \right )^{\ *}_{\ \alpha}
 \vspace*{0.3cm} \\
 \hat{\nu}_{\ n} \ = \ U_{\ m n} \ \nu_{\ m}
\hspace*{0.2cm} ; \hspace*{0.2cm} 
m,n \ = \ 1, \cdots, \ 6
 \vspace*{0.3cm} \\
\mbox{and for k,l = 1,2,3 :}
 \vspace*{0.3cm} \\
\nu_{\ k} \ = \ \left ( \ \overline{U}_{\ (11)} \ \right )_{\ k l} \ \hat{\nu}_{\ l}
\ + \ \left ( \ \overline{U}_{\ (12)} \ \right )_{\ k l} \ \hat{{\cal{N}}}_{\ l}
 \vspace*{0.3cm} \\
{\cal{N}}_{\ k} \ = \ \left ( \ \overline{U}_{\ (21)} \ \right )_{\ k l} \ \nu_{\ l}
\ + \ \left ( \ \overline{U}_{\ (22)} \ \right )_{\ k l} \ \hat{{\cal{N}}}_{\ l}
 \vspace*{0.3cm} \\
\nu^{\ *}_{\ k} \ = \ \left ( \ U_{\ (11)} \ \right )_{\ k l} \ \hat{\nu}^{\ *}_{\ l}
\ + \ \left ( \ U_{\ (12)} \ \right )_{\ k l} \ \hat{{\cal{N}}}^{\ *}_{\ l}
 \vspace*{0.3cm} \\
{\cal{N}}^{\ *}_{\ k} \ = \ \left ( \ U_{\ (21)} \ \right )_{\ k l} \ \nu^{\ *}_{\ l}
\ + \ \left ( \ U_{\ (22)} \ \right )_{\ k l} \ \hat{{\cal{N}}}^{\ *}_{\ l}
\end{array}
\end{equation}

\noindent
{\bf 'low-low' oscillation states and amplitudes}
\vspace*{0.1cm}

\noindent
For 'low-low' oscillations only the 3 by 3 matrices $U_{\ (11)}$ and 
$\overline{U}_{\ (11)}$ are operative.

\noindent
First we consider production of a normalized state (without subscript (0) ) at t = 0 of $nu$ type k :

\begin{equation}
\label{eq:30}
\begin{array}{l} 
\left ( \ U_{\ (11)} \ \right )_{\ kl} \ = \ {\cal{A}}_{\ k l}
\hspace*{0.2cm} ; \hspace*{0.2cm} {\cal{A}} \ = \tau \ U_{\ 0}
 \vspace*{0.3cm} \\
 \left | \ \vec{p} \ ; \ t = 0 \ ; \ k \ \right \rangle_{ (0)} \ = 
\ {\cal{A}}_{\ k l} \ \left | \ \vec{p} \ ; \ m_{\ l} \ ; \ 0 \ \right \rangle
\vspace*{0.3cm} \\
\left | \ \vec{p} \ ; \ t = 0 \ ; \ k \ \right \rangle_{ (0)} \ = 
\ N_{\ k} \ \left | \ \vec{p} \ ; \ t = 0 \ ; \ k \ \right \rangle 
\vspace*{0.3cm} \\
N_{\ k}^{\ 2} \ = \ \left ( \ {\cal{A}} \ {\cal{A}}^{\ \dagger} \ \right )_{\ k k}
\ = \tau^{\ 2}_{\ kk}
\hspace*{0.2cm} \rightarrow \hspace*{0.2cm} 
 \vspace*{0.3cm} \\
\left | \ \vec{p} \ ; \ t = 0 \ ; \ k \ \right \rangle 
\ = \ \varrho_{\ k} \ {\cal{A}}_{\ k l}
\ \left | \ \vec{p} \ ; \ m_{\ l} \ ; \ 0 \ \right \rangle
\hspace*{0.2cm} ; \hspace*{0.2cm} 
\varrho_{\ k} \ = \ 1 \ / \ N_{\ k}
\end{array}
\end{equation}

\noindent
In eq. (\ref{eq:30}) we decomposed the matrix $U_{\ (11)} \ = \ {\cal{A}}$
into the product of a hermitian positive matrix $\tau$ and a unitary 
(3 by 3) matrix $U_{\ 0}$. The deviation of $\tau$ with $0 \ \leq \ \tau \ \leq \ 1$
from unity is of the generic order $m_{\ light}  / M_{\ heavy} \ \sim \ 10^{\ -21}$,
yet it is one of the {\it subtle} properties of neutrino flavor mixing, that this
deviation is at the very origin of the light neutrino masses. This is only so,
if we {\it assume} the generation of these masses through mixing with heavy flavors, as we do here.
\vspace*{0.1cm}

\noindent
{\it subtle things :}
\vspace*{0.1cm}

\begin{description}
\item - the normalization $\varrho_{\ k} \ {\cal{A}}_{\ k l}$ can be enforced by
properly tagging $\nu_{\ k}$ upon production irrespective of decay rates, e.g. for
$\pi^{\ +} \ \rightarrow \ \underline{\mu^{\ +}} \ + \ \nu_{\ \mu} \ + \ n \ \gamma$.

\item - spin states ,  probabilities for helicities

\begin{displaymath}
\begin{array}{lll} 
- 1 & : & 1 \ - \ ( \ 1 \ - \ v \ ) \ / \ 2
 \vspace*{0.3cm} \\
+ 1 & : & ( \ 1 \ - \ v \ ) \ / \ 2 \ \sim  
\ m_{\ \nu}^{\ 2} \ / ( \ 4 \ E^{\ 2} \ )
\end{array}
\end{displaymath}

The precise probability of the 'wrong' helicity state does depend on
the nature of production and decay amplitudes. 

\item - to go from neutrino flavors to antineutrino flavors amounts to
exchange the helicities and to substitute ${\cal{A}} \ \rightarrow \ \overline{{\cal{A}}}$.

\end{description}

\noindent
{\bf Oscillation amplitudes}
\vspace*{0.1cm}

\noindent
We denote the amplitude of transition from t = 0 to t and from initially produced
neutrino flavor k to neutrino flavor k' by $T_{\ k' \ \leftarrow \ k}$, neglecting
the 'wrong' helicity states, and likewise by $AT_{\ k' \ \leftarrow \ k}$ the corresponding 
antineutrino amplitude.

\begin{equation}
\label{eq:31}
\begin{array}{l} 
T_{\ k' \ \leftarrow \ k} \ ( \ t \ ) \ = \ \varrho_{\ k'} \ \varrho_{\ k}
\ \left ( \ {\cal{A}} \ U_{\ diag} \ ( \ t \ ) \ {\cal{A}}^{\ \dagger}
\ \right )_{\ k' k}
 \vspace*{0.3cm} \\
AT_{\ k' \ \leftarrow \ k} \ ( \ t \ ) \ = \ \varrho_{\ k'} \ \varrho_{\ k}
\ \left ( \ \overline{{\cal{A}}} \ U_{\ diag} \ ( \ t \ ) \ \overline{{\cal{A}}}^{\ \dagger}
\ \right )_{\ k' k}
\vspace*{0.3cm} \\
U_{\ diag} \ ( \ t \ ) \ = \ diag 
\ \left ( \ e^{\ -i \ E_{\ 1} \ t} \ , \ e^{\ -i \ E_{\ 2} \ t} \ , \ e^{\ -i \ E_{\ 3} \ t}
\ \right )
 \vspace*{0.3cm} \\
E_{\ j} \ = \ \sqrt{\ p^{\ 2} \ + \ m_{\ j}^{\ 2} \ }
\ \sim p \ + \ \frac{1}{2} \ m_{\ j}^{\ 2} \ / \ p
\end{array}
\end{equation}

\noindent
The amplitudes in eq. (\ref{eq:31}) describe oscillations in vacuo. Matter effects modify 
${\cal{A}}$ in an energy dependent way.

\noindent
CPT invariance is manifest, whereas T invariance requires ${\cal{A}}$ to be real :

\begin{equation}
\label{eq:32}
\begin{array}{lll} 
CPT & : & \left ( \ AT_{\ k \ \leftarrow \ k'} \ ( \ - t \ ) \ \right )^{\ *} \ =
\ T_{\ k' \ \leftarrow \ k} \ ( \ t \ )
\vspace*{0.3cm} \\
T & : & 
\left ( \ T_{\ k \ \leftarrow \ k'} \ ( \ - t \ ) \ \right )^{\ *} \ =
\ T_{\ k' \ \leftarrow \ k} \ ( \ t \ )
\vspace*{0.3cm} \\
 & & \mbox{and} \ T \ \leftrightarrow \ AT
\vspace*{0.3cm} \\
 & & \rightarrow \ {\cal{A}} \ = \ \overline{{\cal{A}}}
\end{array}
\end{equation}

\noindent
{\bf Counting phases}
\vspace*{0.1cm}

\noindent
We generalize again the counting of imaginary parameters or equivalently of complex phase
factors to $3 \ \rightarrow N_{\ h}$ heavy neutrino flavors wit $n \ = \ 3 + \ N_{\ h}$.

\noindent
The number of independent imaginary parameters in the mass matrix ${\cal{M}}$
of the form given in eq. (\ref{eq:16}) is 

\begin{equation}
\label{eq:33}
\begin{array}{lll} 
\# \ \varphi \ ( \ {\cal{M}} \ ) \ = 
\ = \ \frac{1}{2} \ n \ ( \ n \ + \ 1 \ ) \ - \ 6
\end{array}
\end{equation}

\noindent
The number of phases is 15 for 6 neutrino flavors.

\section{Conclusions and outlook}
\vspace*{0.1cm}

\noindent
\begin{description}
\item - The major experimentally accessible features : $m_{\ 1,2,3}$ and $U_{\ 0}$ as
defined in eq. (\ref{eq:30}), i.e. light neutrino mass and mixing
hopefully will establish the specific structure of $\nu \ - \ {\cal{N}}$ dynamics. 

\item - the {\it subtle} and small effects : 1) ${\cal{A}} \ = \ \tau \ U_{\ 0}$ 
with $\tau \ \neq \ 1$  and 2) lepton flavor violation in conjunction with the
observation of 'wrong' neutrino helicities or neutrinoless double $\beta$ decay
are in generic situations expected to be very small. 

\item - CP $\leftrightarrow$ T violating effects in the $\nu \ , \ \overline{\nu}$
leptonic sector are much richer than in the case of quarks and antiquarks.
They are tied to the small masses of the light neutrino flavors. Despite this
'low' energy obstruction it seems to me to be worthwhile to look for these effects
especially at maximally feasible energies, where light - heavy $\nu \ \rightarrow \ {\cal{N}}$
transitions begin to play a role.

\item - "weniges ist mehr $\cdots$".
\end{description}

\subsection*{Acknowledgements}

I thank the organizers of the 'Epiphany 2001' conference in Cracow,
especially Marek Jezabek, for their warm hospitality. 
Special thanks are due to Sonja Kabana for interesting discussions and suggestions.


\end{document}